# The impact of the near-surface region on the interpretation of x-ray absorption spectroscopy

*Lishai Shoham,\* Maria Baskin, Yaron Kauffmann, Anna Zakharova,[†] Teppei Yoshida, Shigeki Miyasaka, Cinthia Piamonteze,\* and Lior Kornblum*

Lishai Shoham, Dr. Maria Baskin, Prof. Lior Kornblum

Andrew and Erna Viterbi Department of Electrical and Computer Engineering, Technion – Israel Institute of Technology, Haifa 3200003, Israel

E-mail: lishai@campus.technion.ac.il

Dr. Yaron Kauffmann

Electron Microscopy Center, Department of Materials Science and Engineering, Technion – Israel Institute of Technology, Haifa 3200003, Israel

Dr. Anna Zakharova, Dr. Cinthia Piamonteze

Swiss Light Source, Paul Scherrer Institute, CH-5232 Villigen PSI, Switzerland

E-mail: cinthia.piamonteze@psi.ch

Prof. Teppei Yoshida

Graduate School of Human and Environmental Studies, Kyoto University, Sakyo-ku, Kyoto 606-8501, Japan

Prof. Shigeki Miyasaka

Department of Physics, Osaka University, Toyonaka, Osaka, 560-0043, Japan.

**Abstract**

Transition metal oxides (TMOs) exhibit a broad spectrum of electronic, magnetic, and optical properties, making them intriguing materials for various technological applications. Soft x-ray absorption spectroscopy (XAS) is widely used to study TMOs, shedding light on their chemical state, electronic structure, orbital polarization, element-specific magnetism, and more. Different XAS acquisition modes feature different information depth regimes in the sample. Here, we employ two XAS acquisition modes, having surface-sensitive versus "bulk" probing depths, on the prototypical TMO $SrVO_3$. We illustrate and elucidate a strong apparent discrepancy between the different modes, emphasizing the impact of the near-surface region on the interpretation of XAS data. These findings highlight the importance of the acquisition mode selection in XAS analysis. Moreover, the results highlight the role of the near-surface region not only in the characterization of TMOs, but also in the design of future nanoscale oxide electronics.

# 1. Introduction

X-ray absorption spectroscopy (XAS) is a widely used technique for studying chemical states, electronic structure, magnetism, and other properties and phenomena.[1–3] For example, in correlated transition-metal oxides (TMOs) at the soft x-ray region, the transition metal L-edge can be used to study the effect of doping and orbital polarization on the valence state.[4,5] In contrast, the oxygen K-edge can be used to study the band structure evolution under different strain states, compositions, etc.[4,6–9] The two most common XAS acquisition techniques are total electron yield (TEY) and total fluorescence yield (TFY). These methods differ in their physical mechanisms and thus probing depths, ranging from typically 5-10 nm for TEY to about 50-100 nm for TFY. An additional acquisition technique is x-ray excited optical luminescence (XEOL), offering an alternative for bulk-sensitive detection for thin films.[10–13] Utilizing multiple techniques simultaneously allows a comprehensive study from different depths of the film.[14,15]

Surfaces of TMOs, especially when exposed to an ambient atmosphere, can be prone to inhomogeneity. The formation of a near-surface region differing from the bulk in symmetry, chemical composition, and other properties can be driven by overoxidation, charged defects, symmetry breaking, and more.[16–20] In addition, changes in the atomic bond length, e.g., as a result of epitaxial strain, can further increase it.[21,22] For example, Sr segregation in perovskite oxides ($ABO_3$) can lead to a Sr-rich surface, reduction or oxidation of the B cation, secondary phase formation, and precipitates at the surface.[16,17,23,24] Understanding the difference between surface and bulk is therefore crucial in most of these systems.

$SrVO_3$ (SVO) is an attractive case study of correlated Mott materials,[25–28] and recently gained traction as an earth-abundant transparent conductive oxide.[29–34] SVO is an early TMO with a cubic perovskite structure and $3d^1$ electronic configuration. Applying biaxial epitaxial strain on SVO films directly affects the V-O bond length, inducing a difference between in-plane (xy) and out-of-plane (z) V-O hybridization.[35] The change in the hybridization partially lifts the V-3$d$ t$_{2g}$ degeneracy, resulting in a preferred occupation as schematically illustrated in Figure 1a. Wu et al. suggested identifying the SVO symmetry breaking by exploiting linearly polarized XAS at the V L$_{2,3}$-edge ($2p \rightarrow 3d$ transitions).[36] The intensity difference between out-of-plane and in-plane linearly polarized XAS gives rise to an x-ray linear dichroism (XLD), selective to the orbital occupancy at a given energy. Therefore, there should be an XLD inversion between tensile and compressive strain stemming from the inverse preferred occupation.[36–39]

Recent XPS studies on SVO thin films revealed a near-surface region as thick as 10 unit cells (u.c.) and composed mainly of $V^{5+}$ states,[40–42] as opposed to the expected $V^{4+}$ in bulk SVO. In light of these results, we decided to revisit the interpretation of the common surface-sensitive TEY mode for the XAS acquisition of TMOs. In this work, we shine a light on the role of the near-surface region in soft-XAS measurements using the test case of the XLD difference measured from strained SVO thin films. The results acquired in XEOL mode clearly illustrate the preferred occupation inversion between the applied tensile and compressive strain. In contrast, the acquisition in TEY mode yields an unexpectedly similar spectrum for tensile, compressive, and unstrained SVO, indicating a significant difference between the two acquisition depths. Our results demonstrate the dramatic contribution of the near-surface region, especially when measuring with surface-sensitive methods, and can be utilized for free surfaces as well as interfaces in a heterostructure.

## 2. Results and Discussion

SVO films (60 u.c.) were grown on two different single crystalline substrates: LaAlO$_3$ (LAO, -1.4% mismatch) and (LaAlO$_3$)$_{0.3}$(Sr$_2$AlTaO$_3$)$_{0.7}$ (LSAT, 0.7%), thus subjecting the SVO to compressive and tensile biaxial strain, respectively. We define the mismatch between the cubic SVO lattice parameter, $a_{SVO}$ = 3.843 Å, and the in-plain cubic/pseudocubic substrate lattice parameter, $a_{sub}$, as 100%·($a_{sub}$-$a_{svo}$)/$a_{sub}$. Details of the structural characterization were reported elsewhere,[27] showing that the films are fully strained to their substrates. Scanning transmission electron microscopy (STEM) taken in high angle annular dark field (HAADF) mode of the strained SVO films are presented in Figures 1b and c, illustrating epitaxial growth and sharp interfaces.

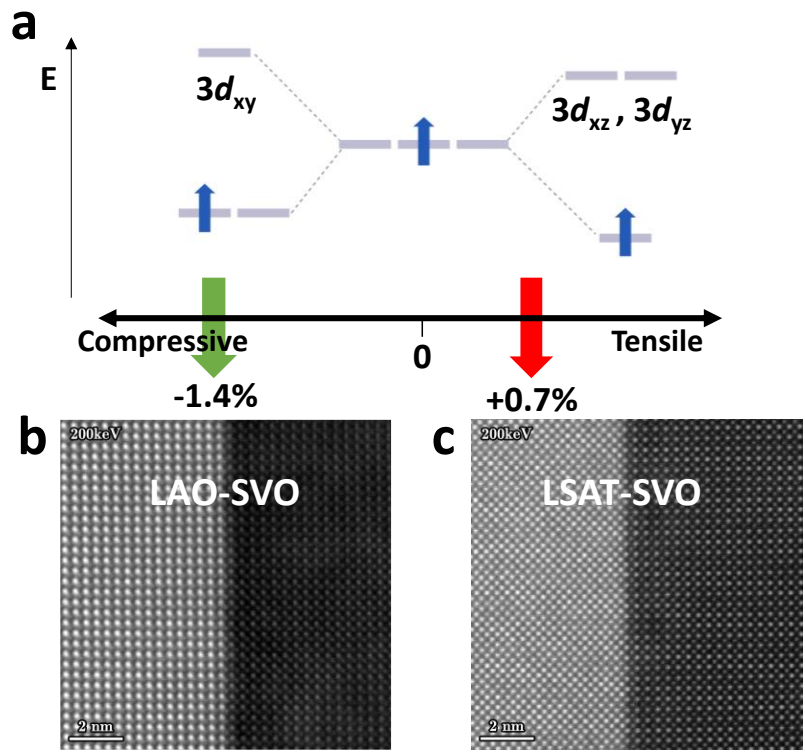

**Figure 1.** a) A schematic of the strain effect on the partial alleviation of the SVO V-3$d$ t$_{2g}$ band. While the t$_{2g}$ states are degenerate in the cubic (unstrained) case, biaxial tensile (SVO-LSAT) or compressive (SVO-LAO) strain stabilize the $d_{xy}$ or the $d_{xz}/d_{yz}$ states, respectively. Given the $d^1$ electronic configuration, the strain results in preferred orbital occupation. Atomic-resolution HAADF-STEM images along the [010] direction of strained SVO films grown on b) LAO and c) LSAT substrates exhibiting coherent growth and high crystallinity of the SVO films.

Tetragonal distortion, applied on the SVO film by the substrate, changes the V-O atomic orbitals overlap, where applying compressive (tensile) strain increases (decreases) the overlap.[27] The effect results in partial removal of the V-3$d$ t$_{2g}$ degeneracy leading to preferred orbital occupation between in-plane and out-of-plane states (Figure 1a). Since XAS at the V L$_{2,3}$-edge probes the V-3$d$ empty states, the intensity difference between out-of-plane ($I_z$) and in-plane ($I_{xy}$) unoccupied states (namely, XLD) provides a measure of the orbital occupation.

### 2.1. XAS by TEY

We start by examining the XAS of the strained and unstrained (single-crystal) 3$d^1$ SVO by employing the surface-sensitive TEY mode. Normalized XAS ($I_z + I_{xy}$) and XLD ($I_z - I_{xy}$) spectra from the V L$_{2,3}$-edge acquired in TEY mode for the SVO system are presented in Figure 2a. The spin-orbit interaction of the V 2$p$ core hole splits the spectrum into L$_3$ (2$p_{3/2}$) at ~ 519 eV and L$_2$ (2$p_{1/2}$) at ~525 eV, whereas the O-K prepeak is in close proximity at ~530 eV (Figure S1). As expected, the SVO TEY-XAS spectra resemble previous TEY-XAS reports for SVO films and crystals.[29,41,43] By comparing the XAS spectra of the unstrained single crystal to the strained films, we can conclude that the strain-induced tetragonal distortion has a negligible effect on the XAS results.

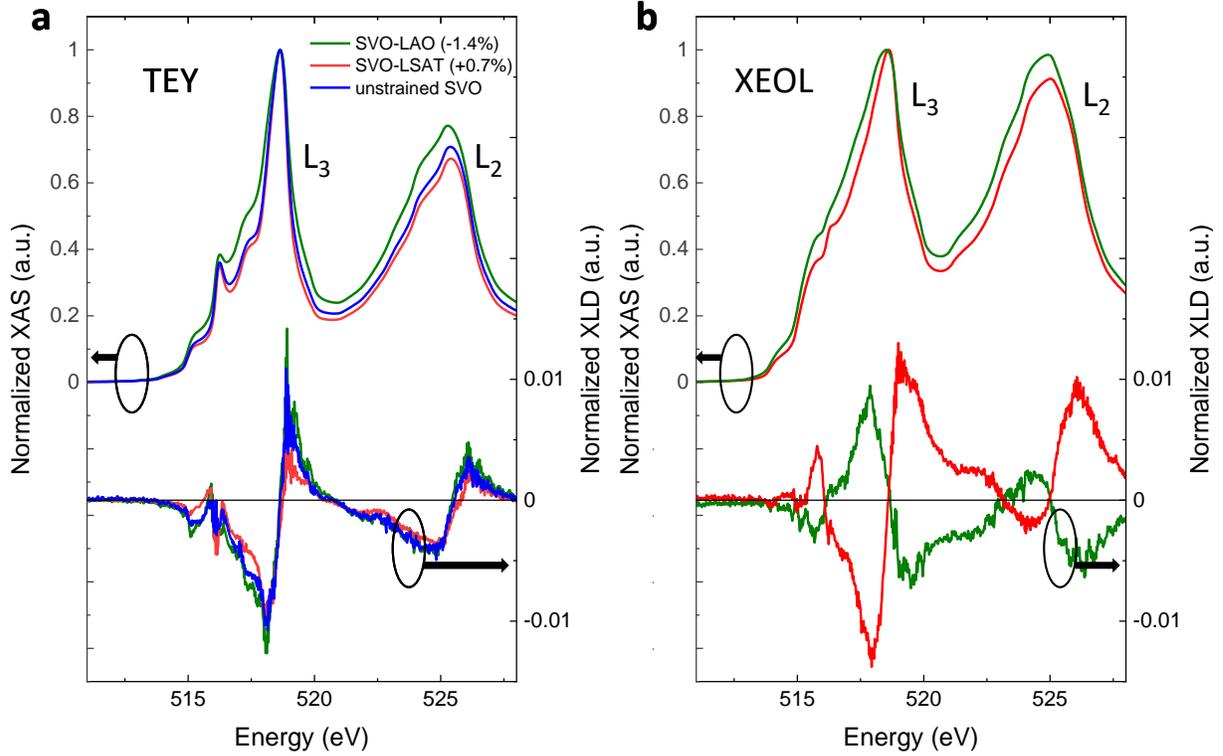

**Figure 2.** Normalized XAS ($I_z + I_{xy}$, top) and XLD ($I_z - I_{xy}$, bottom) spectra at the V $L_{2,3}$-edge from strained and unstrained SVO acquired in a) TEY and b) XEOL modes. After background subtraction, the XAS was normalized to unity at the V $L_3$-edge, and the same factor was used to normalize the XLD spectra. Unstrained SVO refers to an SVO crystal.

We now focus on the TEY-XLD results shown at the bottom of Figure 2a. As schematically illustrated in Figure 1a, unstrained SVO has three-fold degenerate V-3$d$ $t_{2g}$ states, whereas, under tensile (compressive) strain, the in-plane (out-of-plane) states are stabilized. Thus, the measured XLD is expected to be around zero for unstrained (cubic) SVO, and to flip its sign between opposite strain states.[36] However, the results in Figure 2a exhibit nearly identical XLD behavior regardless of the strain state. Moreover, the unstrained SVO presents a similar non-zero XLD behavior, which is completely unexpected due to its cubic structure. We note that few degrees of polishing misalignment are possible with the single crystal; however, the similarity of the spectrum to the thin films suggests that this is not significant for the results. The observations of nearly identical XLD behavior contradict theoretical predictions for SVO,[36] and experimental results on other perovskite TMOs such as nickelates,[37] manganites,[38,44] and ferrites.[39]

## 2.2. XAS by XEOL

Recently, a near-surface region as thick as 10 u.c. was reported in SVO thin films,[40–42] possibly originating from air exposure. This surface region was found to be electrically insulating and to consist of $V^{5+}$ moieties and a secondary Sr chemical state. In light of these works and the TEY-XLD results, we extended our probing depth further into the SVO layer to determine whether the TEY spectra stem mostly from the near-surface region.

Normalized XAS and XLD spectra of the V $L_{2,3}$-edge acquired in XEOL mode for the strained SVO films are presented in Figure 2b. In this detection mode, the optical luminescence created in the substrate by the film's transmitted x-ray is collected using a photodiode located precisely behind the film.[45] For this reason, XEOL was not appropriate for the measurement of the single crystal. The XEOL-XLD spectra clearly illustrate the expected XLD reversal between tensile and compressive strained SVO films.[27] This reversal stems from the strain-dependent tetragonal distortion causing an inversed preferred occupation between tensile and compressive strain. We note that a more common approach for extending the probing depth of TEY is TFY; however, TFY can have several issues, including a large oxygen signal from the substrate (see Supporting Information for details).

We assign the discrepancy between TEY and XEOL results to the presence of the near-surface region and the different probing depths of the two methods. While the surface-sensitive TEY mode probes mainly the near-surface region, thus producing a similar XLD spectrum for the different strained SVO films, the XEOL mode probes the entire thickness ("bulk") of the SVO film, therefore yielding the expected XLD inversion. Moreover, this hypothesis is further strengthened by the non-zero XLD from the unstrained SVO acquired in TEY mode.

The near-surface region phenomenon is not exclusive to SVO, and it was reported for titanates,[14,20] manganites,[18,46,47] and others. We note that the inhomogeneity of the near-surface region can be somewhat negligible for some TMOs. In addition, this inhomogeneity can be advantageous for applications like solid-state catalysis, often becoming the focal point of the analysis. However, for several TMOs, such as SVO, the near-surface region can be substantial and could potentially impede the advancement of devices reliant on ultrathin films and field effect. Hence, we emphasize the potential impact of the near-surface region on the interpretation of XAS results, particularly when these are obtained only via surface-sensitive modes such as TEY.

2.3. EELS Analysis

A complementary analysis of the strained SVO films was done by employing electron energy loss spectroscopy (EELS) at the V $L_{2,3}$-edge, as presented in Figure 3a. The spectra were taken from the middle of the film, as schematically illustrated in the inset of Figure 3a. We note that the reduction of the SVO surface during the lamella preparation prevented EELS analysis of the SVO near-surface region. The EELS results agree with previous EELS analyses done on vanadate films.[48] Both EELS and XAS probe the same electronic transitions but offer different perspectives. While XAS is the only route for obtaining XLD, EELS enables to 'bypass' the near-surface region by probing only at the middle of the film thickness in cross section (we note that the surface of the SVO TEM lamella would also have a near-surface regime, but its portion from the lamella thickness is smaller). The comparison between EELS, TEY, and XEOL for strained SVO films is presented in Figure 3b.

A useful estimate of the V oxidation state, established using EELS, is based on the intensity ratio of the edge peaks, namely the V $L_3/L_2$ ratio.[52–54] Both XAS and EELS studies on vanadate crystals reported a ratio of $L_3/L_2 \sim 1$ for $V^{4+}$ and $L_3/L_2 > 1$ for $V^{5+}$ or $V^{3+}$.[48,55] We report on an $L_3/L_2$ ratio for tensely (compressively) strained SVO of 1.06 (1.05) using EELS, 0.91 (0.98) using XEOL, and 0.67 (0.77) using TEY. We suggest that the discrepancy between ratios taken using the same acquisition technic stems from a small thickness variation of the near-surface region. Considering our previous XPS study,[40] we assign the XAS-TEY spectrum mainly to a $V^{5+}$ phase present in the near-surface region. In contrast, the XEOL-XAS spectrum stems from a combination of a majority of $V^{4+}$ from bulk SVO and a minority of $V^{5+}$ from the near-surface region. We note that the XEOL $L_3/L_2$ ratio is approaching that obtained by EELS, signifying the small contribution of the near-surface region when the spectrum is acquired from the entire film.

It was previously shown that strain-dependent oxygen stoichiometry could significantly affect the measured physical properties, e.g., inducing a metal to insulator transition;[49,50] and it can affect the orbital polarization.[51] However, in this work, we can rule out strain-induced oxygen vacancies due to the similarity of the EELS V $L_{2,3}$-edge between the tensely and compressively strained SVO films.

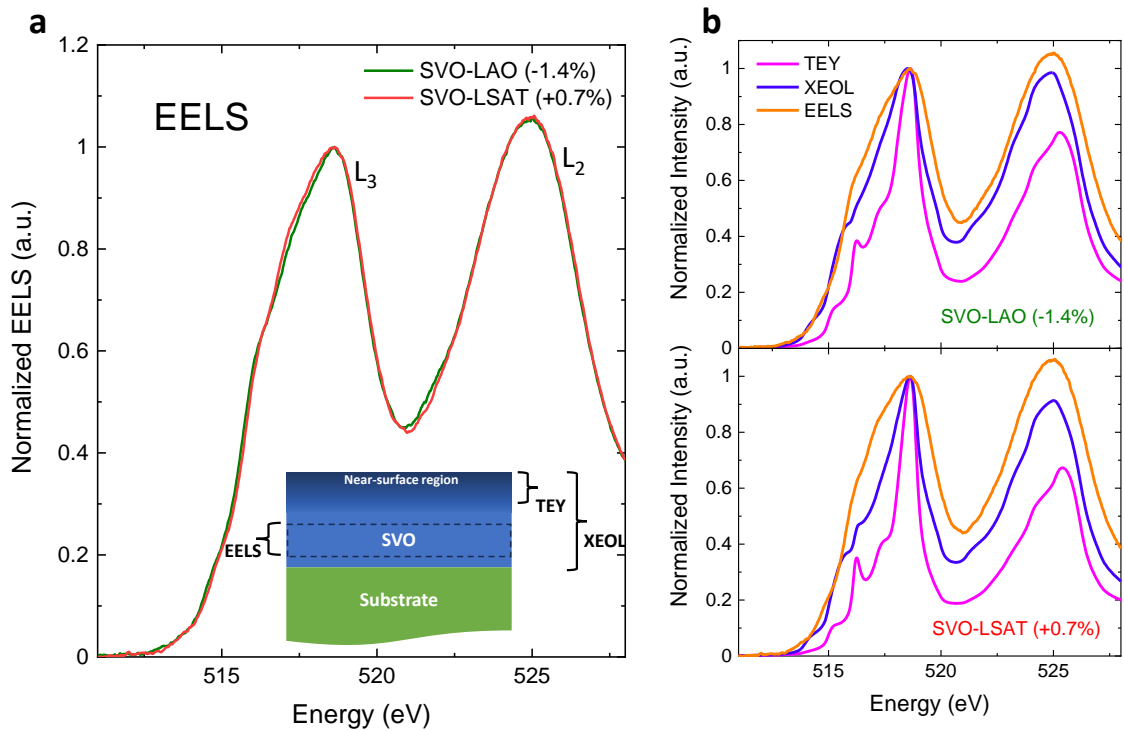

**Figure 3.** a) Normalized EEL spectra of the V $L_{2,3}$ edge measured from the strained SVO films. Each spectrum was generated by averaging several layers from the middle of the film. After background subtraction, the spectrum was normalized to unity at V $L_3$-edge. The insert shows a schematic illustration of the probing area of each acquisition method. b) A comparison of the normalized intensity of the V $L_{2,3}$ edge acquired in TEY, XEOL, and EELS modes for compressive and tensile strained SVO (SVO-LAO and SVO-LSAT, respectively) is presented in the upper and lower panels, correspondingly.

2.4. The Near-Surface Region

The XLD results acquired in TEY are identical regardless of the strain states, and compared to the XEOL results, the SVO in TEY appears to be all tensely strained (Figure 4). Ishida et al. used *ab-initio* calculations to show that free SVO surfaces feature surface relaxation, resulting in orbital polarization.[56] Taken on their own, it is appealing to attribute the TEY results to such surface relaxation. However, these calculations considered ideal SVO surfaces ($V^{4+}$), whereas in the present case, we studied air-exposed SVO, which was shown to have a near-surface region,[40] dominated by $V^{5+}$. Nonetheless, despite the chemical and microstructural differences between these cases, it is possible that the near-surface region in our films still experiences some surface relaxation which acts in a similar manner to the theoretical prediction, inducing the observed orbital polarization in TEY-XLD.

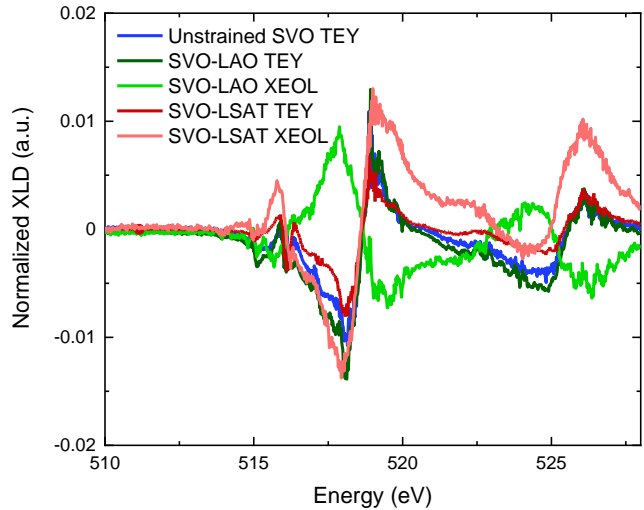

**Figure 4.** A comparison of the normalized XLD spectra at V $L_{2,3}$-edge for strained and unstrained SVO acquired in TEY and XEOL modes. We measured a single crystal for unstrained SVO, and thus the spectrum cannot be acquired via XEOL mode.

The presence of $V^{5+}$ and $Sr^{2+}$ at the near-surface region suggests that the possible composition of the secondary phase is $Sr_3V_2O_{8+\delta}$.[57–60] In addition, when the Sr concentration at the secondary surface phase is higher than its solubility limit, Sr-rich precipitates are formed on the surface, as showed for SVO films,[59] and other perovskite TMOs such as SrTiO$_3$.[61] If there are crystalline phases in the near-surface region, we consider their quantity to be too small to be detected using x-ray diffraction (see Figure S2 for wide range XRD scans). Further work is necessary to fully clarify the composition and structure of the SVO near-surface region.

## 3. Conclusions

In this work, we pinpoint the potential effect of TMOs near-surface region on surface-sensitive methods via soft-XAS analysis of strained SVO thin films. Utilizing the SVO inverse symmetry breaking under biaxial compressive and tensile strain, we distinguish between the near-surface region and the "bulk" via the different probing depths of TEY and XEOL. While XEOL-XLD spectra, capable of probing the entire film thickness, present the expected signal inversion between tensile and compressive strain, the surface-sensitive TEY-XLD spectra show similar behavior for all SVO samples. Using a complementary EELS analysis and a previous XPS study, we can conclude that the near-surface region is composed mainly of V$^{5+}$ secondary phase, whereas the SVO "bulk" matches the V$^{4+}$ as expected. We note that TEY is a well-used method, highly suitable for investigating the surface/interface of the material or when the TMO near-surface region is similar enough to the bulk. Nevertheless, our findings emphasize the importance of understanding the role and magnitude of the near-surface region for all applications: bare films, heterostructures, and especially ultrathin films. The consequences of this effect and its possible analysis artifacts are essential for fundamental studies and device applications.

## 4. Experimental Section

SVO films (60 u.c.) were epitaxially grown on as received (001)-oriented (cubic\pseudocubic) LSAT and LAO 5x5 mm$^2$ single-side polished substrates from CrysTec GmbH. The growth was done in a customized Veeco GenXplor reactive oxide molecular beam epitaxy (MBE) instrument with a base pressure of ~5x10$^{-10}$ Torr. The substrate growth temperature was 1000 °C, as indicated by a thermocouple in proximity to a Mo back plate in contact with the substrate during growth. The Sr and V atomic fluxes were independently calibrated in vacuum prior to the growth using a quartz crystal microbalance (QCM) to 1 u.c./min growth rate. The growth parameters were based on our earlier work,[34] and adapted to a shuttered growth method.[40,62] The oxygen pressure during growth was kept at a non-line-of-sight ionization gauge reading of ~6×10$^{-7}$ Torr for the duration of the growth and the cooling down. An SVO single crystal was grown by the floating-zone method, as described elsewhere.[63]

The XAS experiment was performed at the XTerme beamline of the Swiss Light Source (SLS).[64] XAS spectra were obtained at the V L$_{2,3}$-edge using two linear polarizations of the incident synchrotron radiation: i) parallel to the film's xy plane (E$_{xy}$), and ii) at 30° from the film's normal (E$_z$), which are considered to be roughly proportional to the film's in-plane and out-of-plane directions, respectively. All measurements were conducted at room temperature. Data acquisition was done via two detection modes: TEY and XEOL. The TEY was acquired by measuring the photocurrent created by the absorbed x-ray, where the XEOL was obtained using a photodiode located at the back of the sample. Moreover, to obtain reliable XLD spectra by XEOL, we used the monochromator setting with c$_{ff}$=1.5 in order to drastically reduce high harmonic contamination.[64] See Supporting Information for more details.

STEM was used to acquire atomic resolution HAADF images and EELS spectra. The lamellas were prepared using a Thermofisher Scientific Helios Nanolab G3 Dual-beam focused ion beam. Before the experiment, the surface of the samples was cleaned by a plasma treatment using a 1020 Fischione Ar-O plasma cleaner. The STEM imaging was done using a monochromated and double corrected Thermofisher Titan Themis G$^2$ 60-300 operated at 200 KeV and equipped with a Fischione HAADF-STEM detector and a Gatan 965 post column EELS detector. The zero-loss peak full width at half-maximum for the energy dispersion used to acquire the data was found to be ~0.25 eV.

## Acknowledgments

This work was funded by the Israel Science Foundation (grant No. 1351/21). We acknowledge partial support in sample preparation and characterization from the Russell Berrie Nanotechnology Institute (RBNI) and the Technion's Micro-Nano Fabrication & Printing Unit (MNF&PU). The authors thank Dr. Larisa Popilevsky (Technion) for TEM samples preparation, and Mr. Brajagopal Das for his critical reading of the manuscript. A.Z. acknowledges the financial support by the Swiss National Science Foundation (SNSF) under Project No. 200021_169467.

Supporting Information

# The impact of the near-surface region on the interpretation of x-ray absorption spectroscopy

*Lishai Shoham, Maria Baskin, Yaron Kauffmann, Anna Zakharova, Teppei Yoshida, Shigeki Miyasaka, Cinthia Piamonteze, and Lior Kornblum*

1. **Oxygen K-edge**

Normalized intensity for V $L_{2,3}$-edge and O K-edge of strained and unstrained $SrVO_3$ (SVO) are presented in Figure S1. The spectra were acquired using x-ray absorption spectroscopy (XAS) in total electron yield (TEY) and x-ray excited optical luminescence (XEOL), and scanning transmission electron microscopy electron energy loss spectroscopy (STEM-EELS). The TEY-XAS spectrum, as elaborated in the main paper, is composed mainly of the SVO near-surface region. In contrast, the XEOL-XAS spectrum stems from the entire thickness of the film, and includes a contribution from the oxygen present in the oxide substrate. This has small effect on the V $L_{2,3}$-edge, but it strongly affects the O K-edge. The STEM-EELS spectrum illustrates the SVO V L-edge and O K-edge with a minor contribution from the near-surface region, as elaborated in the main text.

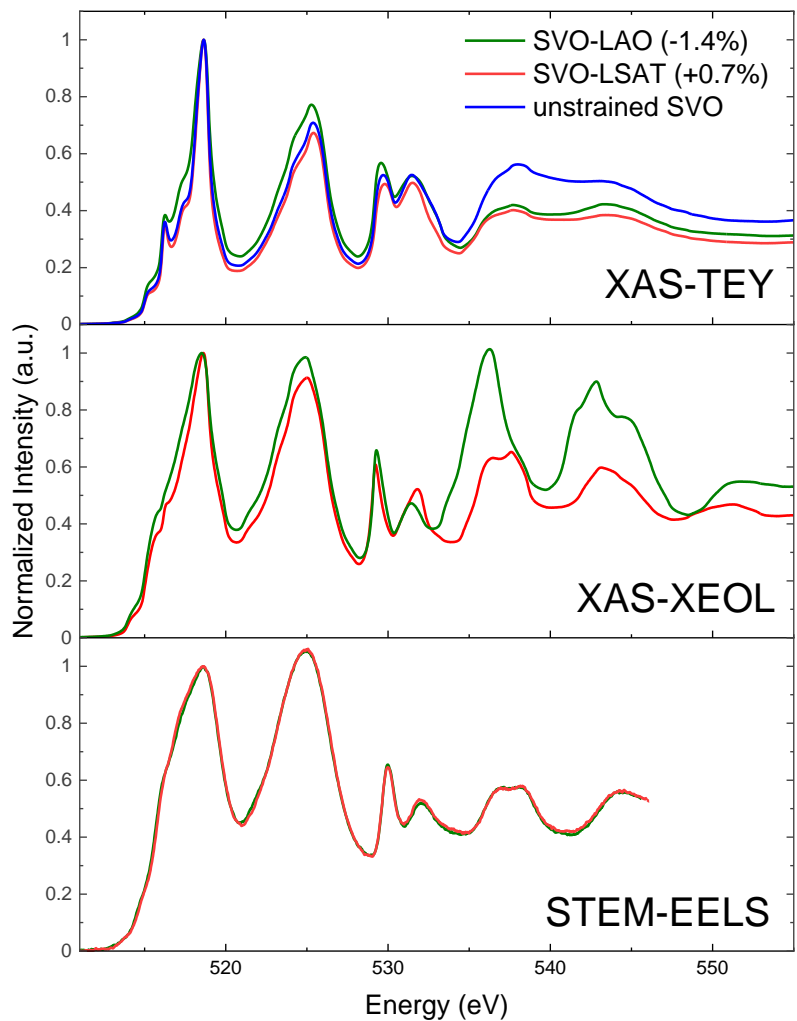

**Figure S1.** Comparison of the normalized V $L_{2,3}$-edge and O K-edge for tensile and compressive strained SVO films (SVO-LSAT and SVO-LAO, respectively) and unstrained (single-crystal) SVO. The acquisition was made using XAS in TEY and XEOL modes, and STEM-EELS. After background subtraction, the spectrum was normalized to unity at V $L_3$-edge.

## 2. X-ray Diffraction (XRD)

XRD scans for the different strained SVO films are presented in Figure S2. The wide scans illustrate the high crystallinity of the SVO films and the absence of epitaxial foreign phases.

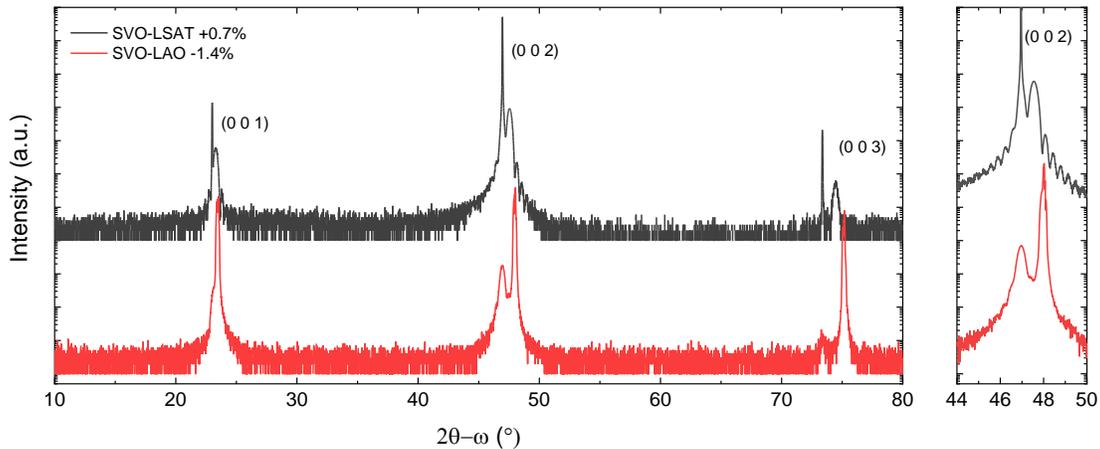

**Figure S2.** High-resolution 2θ-ω scans around (001), (002), and (003) peaks of LSAT (black) and LAO (red) substrates. The scans were acquired using a Rigaku SmartLab with a 2-bounce incident monochromator.

## 3. Total Fluorescence Yield (TFY)

A common approach for extending the probing depth of TEY is TFY. However, in some cases, TFY can have several issues: (i) it can lead to self-absorption for concentrated samples, which makes the $L_3$-edge smaller than the $L_2$. (ii) The TFY is not always equivalent to the x-ray absorption spectra.[1] And, finally, (iii) depending on the edge investigated and the strength of background fluorescence, it can lead to negative peaks in the measured absorption.[2] This does not apply to all samples measured by TFY. For our films, the large O TFY signal from the film and substrate can be observed in Figure S3 to interfere with the V-$L_{2,3}$ spectral region.

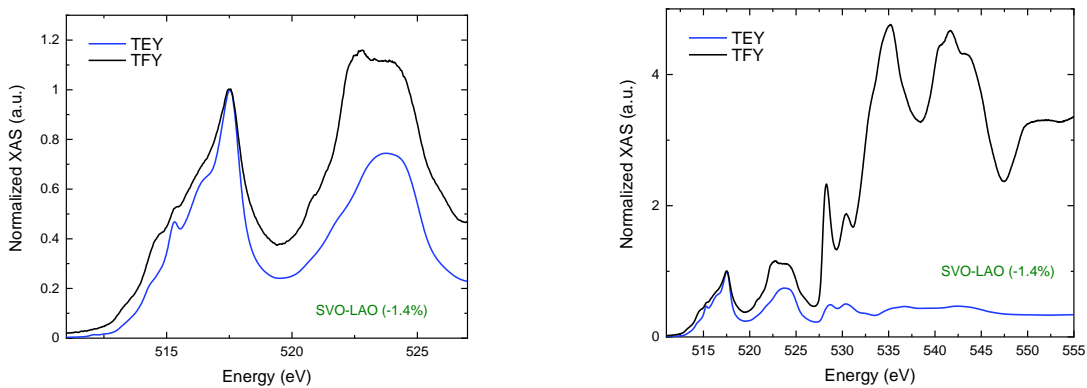

**Figure S3.** Comparison of the V $L_{2,3}$-edge and O K-edge for compressively strained SVO (SVO-LAO) acquired in TEY and TFY modes.

## 4. Reduced High Harmonic Contamination For XEOL

To obtain a reliable XLD spectrum by XEOL, we used the monochromator setting with $c_{ff}=1.5$ in order to drastically reduce high harmonic contamination.[3] Using more standard values of $c_{ff}$ (like 2.25 or 5.0), we always had a different spectrum that looked more like the XAS spectrum. We believe this comes from the fact that the flux for linearly polarized x-rays in the second harmonic is drastically different if the polarization is horizontal or vertical. However, more measurements are needed to be sure.